\documentclass[aps,prd,notitlepage,showpacs,nofootinbib,superscriptaddress]{revtex4-1}
\usepackage{graphicx}
\usepackage[utf8]{inputenc} 

\usepackage[titletoc]{appendix}

\usepackage{amsmath}
\usepackage{amssymb}
\usepackage{float}
\usepackage{comment}
\usepackage{slashed}
\usepackage[normalem]{ulem}
\usepackage{subcaption}
\usepackage{caption}
\usepackage{subcaption}
\usepackage{amsmath}
\usepackage{amssymb}

\usepackage[usenames,dvipsnames]{color}
\usepackage[colorinlistoftodos]{todonotes}
\usepackage[colorlinks=true,citecolor=darkred,urlcolor=darkred, pdfborder={0 0 0}]{hyperref}
\usepackage[normalem]{ulem}

\definecolor{darkred}{rgb}{0.6,0,0}
\usepackage[colorinlistoftodos]{todonotes}

\definecolor{linkcolor}{rgb}{0,0,0.5}
\newcommand {\ignore}[1]{}

\usepackage{soul}

\definecolor{mightnightblue}{RGB}{25,25,112}

\definecolor{brown}{rgb}{0.59, 0.29, 0.0}

\newcommand {\black} {\color{black}}

\begin{document}

%\preprint{IPM/P-2012/009}
%\vspace*{3mm}

\title{\boldmath \color{BrickRed} Scotogenic dark matter from gauged $B-L$}

\author{Yadir Garnica}\email{ya.garnicagarzon@ugto.mx}
\affiliation{Departamento de F\'isica, DCI, Campus Le\'on, Universidad de
  Guanajuato, Loma del Bosque 103, Lomas del Campestre C.P. 37150, Le\'on, Guanajuato, M\'exico}
  
\author{América Morales}\email{america@fisica.ugto.mx }
\affiliation{Departamento de F\'isica, DCI, Campus Le\'on, Universidad de
  Guanajuato, Loma del Bosque 103, Lomas del Campestre C.P. 37150, Le\'on, Guanajuato, M\'exico}  

\author{Carlos A. Vaquera-Araujo}\email{vaquera@fisica.ugto.mx}
\affiliation{Consejo Nacional de Humanidades, Ciencias y Tecnolog\'ias, Av. Insurgentes Sur 1582. Colonia Cr\'edito Constructor, Del. Benito Ju\'arez, C.P. 03940, Ciudad de M\'exico, M\'exico}
\affiliation{Departamento de F\'isica, DCI, Campus Le\'on, Universidad de
  Guanajuato, Loma del Bosque 103, Lomas del Campestre C.P. 37150, Le\'on, Guanajuato, M\'exico}
\affiliation{Dual CP Institute of High Energy Physics, C.P. 28045, Colima, M\'exico}

%%%%%%%%%%%%
\begin{abstract}
\vspace{0.5cm}
We propose a $U(1)_{B-L}$ gauge extension to the SM, in which the dark sector is stabilized through a matter parity symmetry preserved
after spontaneous symmetry breaking. The fermion spectrum includes three neutral right-handed fields 
with $B-L$ charges $(-4,-4, 5)$, that make the model free of gauge anomalies. Two of these neutral fermion fields serve as mediators in a scotogenic mechanism for light-active Majorana neutrino masses. The corresponding 
neutrino mass matrix has rank 2, predicting a massless state and a lower bound for neutrinoless double beta decay. Regions in the parameter space consistent with dark matter relic 
abundance are accomplished by the lightest neutral mediator.
\end{abstract}
%%%%%%%%%%%%

\maketitle
\noindent

%%%%%%%%%%%%%%%%%%%%%%%%
\section{Introduction}
\label{Sect:intro}
%%%%%%%%%%%%%%%%%%%%%%%%
Despite the significant success of the Standard Model (SM) in describing the strong and electroweak (EW) interactions, there remain fundamental issues that the SM cannot directly address, such as the nature and origin of neutrino masses. Another critical limitation of the SM is its inability to provide a viable and stable Dark Matter (DM) candidate. Addressing these challenges requires invoking physics beyond the SM. One particularly compelling scenario is the scotogenic model, where the mechanism for generating neutrino masses is mediated by a DM candidate \cite{Tao:1996vb,Ma:2006km}. In this model, the active neutrino masses are generated at one loop, and the lightest electrically neutral mediator in the loop can be identified as a Weakly Interacting Massive Particle (WIMP), stabilized via a global $\mathbb{Z}_2$ symmetry, offering a unified explanation for these cosmological and particle physics mysteries.

An attractive alternative to the original scotogenic model is the possibility that the symmetry stabilizing DM is not as an {\emph ad hoc} imposition but emerges from a symmetry closely related to neutrinos. The SM exhibits an accidental global $\mathrm{U}(1)_{B-L}$ symmetry, where $B$ denotes baryon number and $L$ stands for lepton number. This symmetry cannot be naturally promoted to a local symmetry due to anomalies. To cancel these anomalies, the fermion content of the model needs to be expanded. The simplest approach is to introduce three right-handed neutrinos with $B-L = -1$. However, another minimal choice is to introduce three electrically neutral right-handed fermions $(N_1, N_2, X)$ with $B-L$ charge assignments $(-4, -4, 5)$ \cite{Montero:2007cd}. In this work, we adopt the latter choice as our starting point. In our model, the gauged $B-L$ symmetry is spontaneously broken by two units, resulting in a residual local $\mathbb{Z}_2$ symmetry after symmetry breaking. This $\mathbb{Z}_2$ symmetry can be identified as matter parity $M_P = (-1)^{3(B-L) + 2s}$, ensuring the stability of the lightest neutral $M_P$-odd particle and potentially providing a viable WIMP DM candidate.

The objective of this paper is to explore the use of $N_1$ and $N_2$ as mediators in a scotogenic Majorana mass generation mechanism for neutrinos.  Notice that this is a minimal model in the sense that the introduction of those fields is one of the minimal choices of extra fermion fields needed to obtain an anomaly free gauge theory based on the $B-L$ symmetry.  Since there are only two dark fermion mediators in the scotogenic mechanism, the resulting light active neutrino mass matrix has rank 2, predicting one massless neutrino and setting a lower bound for neutrinoless double beta decay.

The paper is organized as follows: In Section II we present the matter content of the model and their transformation properties under the gauge symmetries. Section III is devoted to the identification of the physical scalars and gauge bosons of the model. The scotogenic mechanism for light active neutrino masses is presented in Section IV and the analysis of neutrinoless double beta decay is contained in Section V. We study the viability of a WIMP DM candidate identified with the lightest neutral scalar in the dark sector in section VI, and we summarize our results in Section VII.

%%%%%%%%%%%%%%%%%%%%%%%%%%%%%%%
\section{The Model}
\label{model}
%%%%%%%%%%%%%%%%%%%%%%%%%%%%%%

\begin{table}[ht]
\centering
\begin{tabular}{|c|c|c|c|c|c|c|}
\hline
\hspace{0.2cm} Field \hspace{0.2cm} & \hspace{0.2cm}SU(3)$_c$ \hspace{0.2cm}
& \hspace{0.2cm} SU(2)$_L$ \hspace{0.2cm} & \hspace{0.2cm}U(1)$_Y$ \hspace{%
0.2cm} & \hspace{0.2cm}U(1)$_{B-L}$ \hspace{0.2cm} &\hspace{0.2cm} $M_P$ \hspace{0.2cm} \\ 
\hline\hline
$Q_{a L}$ & \textbf{3} & $\textbf{2}  $ & $\frac{1}{6}$ & $\frac{1}{3}$ &  $+$ \\ 
$u_{aR}$ & \textbf{3} & \textbf{1} & $\frac{2}{3}$ & $\frac{1}{3}$  & $+$ \\ 
$d_{aR}$ & \textbf{3} & \textbf{1} & $-\frac{1}{3}$ & $\frac{1}{3}$ & $+$ \\
$L_{aL}$ & \textbf{1} & \textbf{2} & $-\frac{1}{2}$ & $-1$  & $+$ \\ 
$e_{a R}$ & \textbf{1} & \textbf{1} & $-1$ & $-1$  & $+$ \\ \hline
$N_{i R}$ & \textbf{1} & \textbf{1} & $0$ & $-4$  & $-$ \\ 
$X_{R}$ & \textbf{1} & \textbf{1} & $0$ & $5$  & $+$\\
$F_{L,R}$ & \textbf{1} & \textbf{1} & $0$ & $3$  & $+$ \\ \hline\hline
$H$ & \textbf{1} & \textbf{2} & $\frac{1}{2}$ & $0$  & $+$ \\ 
\hline
$\phi_2$ & \textbf{1} & \textbf{1} & $0$ & $2$  & $+$ \\
$\phi_6$ & \textbf{1} & \textbf{1} & $0$ & $6$ & $+$ \\
$\phi_8$ & \textbf{1} & \textbf{1} & $0$ & $8$  & $+$ \\
$\eta$ & \textbf{1} & \textbf{2} & $\frac{1}{2}$ & $-3$  & $-$ \\ 
$\sigma$ & \textbf{1} & \textbf{1} & $0$ & $-3$  & $-$ \\
 \hline
\end{tabular}%
\caption{Particle content of the model ($a=1,2,3$ and $i =1,2$ represent
generation indices).}
\label{tab3211} 
\end{table}

The model is a SM extension based on the $\mathrm{SU}(3)\otimes \mathrm{SU}(2)_W\otimes\mathrm{U}(1)_Y\otimes \mathrm{U}(1)_{B-L}$ gauge symmetry.
The particle content of the model is shown in Table \ref{tab3211}. Beside the SM fields, the model includes three right-handed fermions $(N_1,N_2,X)$, a vector-like fermion $F$, three active scalar singlets $\phi_\alpha$, $\alpha\in\{2,6,8\}$, a dark scalar inert doublet $\eta$ and a dark singlet $\sigma$.  As will be shown below, after Spontaneous Symmetry Breaking (SSB), $B-L$ is broken by two units, leaving a remnant unbroken discrete symmetry known as  matter parity $M_P=(-1)^{3(B-L)+2s}$.
The dark fermions $N_i$ together with the new scalars serve as mediators in a scotogenic neutrino mass generation mechanism, and the lightest dark particle is stabilized by the discrete gauge symmetry $M_P$. It can be identified as a WIMP DM candidate.  The vector-like fermion $F$ is needed to induce a mass for the $X$ fermion\footnote{Notice that in order to give masses to $N_i$ and $X$, the obvious choice is to include only two $M_P$-even scalar singlets transforming as $\phi_8\sim(\mathbf{1},\mathbf{1},0,8)$ and $\phi_{10}\sim(\mathbf{1},\mathbf{1},0,10)$. However, this simple setup leads to the presence of an unobserved Nambu-Goldstone boson.}. 

%%%%%%%%%%%%%%%%%%%%%%%%%%%%%%%%%%%%
\section{Boson Spectrum}
\label{sec:bosons}
%%%%%%%%%%%%%%%%%%%%%%%%%%%%%%%%%%%%
The gauged $B-L$ symmetry is spontaneously broken by two units as the singlet scalar $\phi_2$ develops a vacuum expectation value (VEV). The most general VEV alignment for the scalar fields compatible with the preservation of $M_P$ symmetry is
\begin{equation}
H=\left(
\begin{array}{c}
 G^{+} \\
 \frac{v+s_1+i a_1}{\sqrt{2}} \\
\end{array}
\right),\quad
\eta=\left(
\begin{array}{c}
 \eta^{+} \\
 \frac{s'_1+i a'_1}{\sqrt{2}} \\
\end{array}
\right),\quad 
\phi_\alpha= \frac{v_\alpha+s_\alpha+i a_\alpha}{\sqrt{2}} ,\quad
\sigma= \frac{s'_2+i a'_2}{\sqrt{2}}.
\end{equation}
The scalar potential invariant under the symmetries of the model is
\begin{equation}\label{Pot}
\begin{split}
V =& \mu_1^2H^\dagger H+\mu_2^2\phi_2^{\star}\phi_2+\mu_6^2\phi_6^{\star}\phi_6+\mu_8^2\phi_8^{\star}\phi_8+\mu_3^2\eta^{\dagger}\eta+\mu_4^2\sigma^{\star}\sigma\\&+\lambda_1(H^\dagger H)^2+\lambda_2(\eta^{\dagger}\eta)^2+\lambda_{3}(H^\dagger H)(\eta^{\dagger}\eta)+\lambda_{4}(H^\dagger \eta)(\eta^{\dagger}H)\\
&+ \lambda_{5}(H^\dagger H)(\phi_2^{\star}\phi_2)+\lambda_{6}(H^\dagger H)(\phi_6^{\star}\phi_6) +\lambda_{7}(H^\dagger H)(\phi_8^{\star}\phi_8)+\lambda_{8}(H^\dagger H)(\sigma^{\star}\sigma) \\
&+\lambda_{9}(\eta^{\dagger}\eta)(\phi_2^{\star}\phi_2)+ \lambda_{10}(\eta^{\dagger}\eta)(\phi_6^{\star}\phi_6)+\lambda_{11}(\eta^{\dagger}\eta)(\phi_8^{\star}\phi_8)+\lambda_{12}(\eta^{\dagger}\eta)(\sigma^{\star}\sigma)\\
&+\lambda_{13}(\phi_2^{\star}\phi_2)^2 +\lambda_{14}(\phi_6^{\star}\phi_6)^2+\lambda_{15}(\phi_8^{\star}\phi_8)^2+\lambda_{16}(\sigma^{\star}\sigma)^2\\
&+ \lambda_{17}(\phi_2^{\star}\phi_2)(\sigma^{\star}\sigma)+ \lambda_{18}(\phi_6^{\star}\phi_6)(\sigma^{\star}\sigma)+\lambda_{19}(\phi_8^{\star}\phi_8)(\sigma^{\star}\sigma)\\
&+\lambda_{20}(\phi_2^{\star}\phi_2)(\phi_6^{\star}\phi_6)+\lambda_{21}(\phi_2^{\star}\phi_2)(\phi_8^{\star}\phi_8)+\lambda_{22}(\phi_6^{\star}\phi_6)(\phi_8^{\star}\phi_8)\\
 &+\left[\lambda_{23}\phi_6^{\star}\phi_2\phi_2\phi_2+\lambda_{24}\phi_2^{\star}\phi_8\sigma\sigma+\lambda_{25}(H^\dagger \eta)\phi_6\sigma+\mathrm{h.c.}\right]\\
 &+\left[\frac{\mu_{t}}{\sqrt{2}}(\eta^\dagger H)\sigma+ \frac{\mu_{s}}{\sqrt{2}} \phi_6 \sigma \sigma - \frac{\mu_{u}}{\sqrt{2}} \phi_8^{\star}\phi_6\phi_2 +\mathrm{h.c.}\right].
\end{split}
\end{equation}
In this work, we assume that the hierarchy between the electroweak breaking scale $v$ and the $B-L$ breaking scales $v_\alpha$, $\alpha\in\{2,6,8\}$ is given by $v\ll v_\alpha$, and for simplicity, we will suppose that all the parameters of the potential are real.
The minimization conditions of the scalar potential are then
\begin{eqnarray}
\mu^2_1 &=& \frac{1}{2} \left(-\lambda _5 v_2^2-\lambda _6 v_6^2-\lambda _7 v_8^2-2 \lambda _1 v^2\right),\nonumber\\
\mu^2_2 &=& \frac{v_6 v_8 \mu_u -v_2 \left( 2 \lambda_{13} v_2^2+\lambda_{20}v_6^2+\lambda_{21} v_8^2+3 \lambda_{23}v_2 v_6+\lambda_5 v^2 \right)}{2 v_2},\nonumber\\
\mu^2_6 &=& -\frac{\lambda _{23} v_2^3-v_8 v_2 \mu _u+v_6 \left(\lambda _{20} v_2^2+2 \lambda _{14} v_6^2+\lambda _{22} v_8^2+\lambda_6 v^2\right)}{2 v_6},\nonumber\\
\mu^2_8 &=& \frac{1}{2} \left(-\lambda _{21} v_2^2-\lambda _{22} v_6^2-2 \lambda _{15} v_8^2+\frac{v_2 v_6 \mu _u}{v_8}-\lambda _7 v^2 \right).
\end{eqnarray}
\subsection{Charged scalars}
From the analysis of the scalar potential, we find that the charged scalars $G^{\pm}$ are identified with the goldstone bosons absorbed by the charged gauge fields $W^{\pm}_\mu$  and remain unmixed with the physical dark charged scalars  $\eta^{\pm}$, which in turn acquire squared masses
\begin{equation}
m_\eta^{2}=\frac{1}{2} \left( v^2 \lambda_3 + v_2^2 \lambda_9 + v_6^2 \lambda_{10} + v_8^2 \lambda_{11} \right).
\end{equation}

\subsection{CP-odd scalars}
In the neutral  CP-odd sector, the squared mass matrix for the dark scalars in the basis $(a^{\prime}_1,a^{\prime}_2)$ is
\begin{equation}
M_1^2=\frac{1}{2}\left(
\begin{array}{cc}
 \left(\lambda _3+\lambda _4\right) v^2+\lambda _9 v _2^2+\lambda _{10} v_6^2+\lambda _{11} v_8^2+2\mu^2_3 &  v \left(\mu _t-\lambda _{25} v_6\right) \\
 v \left(\mu _t-\lambda _{25} v_6\right) & \lambda _8 v^2+\lambda _{17} v_2^2+\lambda _{18}v_6^2+\lambda _{19} v_8^2-2 \lambda _{24} v_2 v_8-2 v_6 \mu _s+ 2\mu^2_4 \\
\end{array}
\right).
\label{eq:ap-matrix}
\end{equation}
Eq. \eqref{eq:ap-matrix} is diagonalized by an orthogonal transformation
\begin{equation}
  \left(\begin{array}{c}
    \tilde{\varphi}_1\\
    \tilde{\varphi}_2
  \end{array}
  \right)
  =U^a
  \left(\begin{array}{c}
    a_1'\\
    a_2'
  \end{array}\right)
  =\left(\begin{array}{cc}
    \cos\theta_a& \sin\theta_a\\
    -\sin\theta_a& \cos\theta_a
  \end{array}\right)
  \left(\begin{array}{c}
    a_1'\\
    a_2'
  \end{array}\right)\, ,
\label{eq:odd_orthogonal}
\end{equation}
where the mixing angle is determined by
\begin{equation}
  \tan{2\theta_a}=\frac{2v(\mu_t-\lambda_{25}v_6)}{v^2(\lambda_3+\lambda_4-\lambda_8)+v_8^2(\lambda_{11}-\lambda_{19})+v_2\left(\lambda_9 v_2-\lambda_{17}v_2+2\lambda_{24}v_8\right)+v_6\left(\lambda_{10}v_6-\lambda_{18}v_6+2\mu_s\right)+2(\mu_3^2-\mu_4^2)}\, .
  \label{eq:t2delta}
\end{equation}
The mass eigenvalues associated with the real physical scalars $\widetilde{\varphi}_1$ and  $\widetilde{\varphi}_2$ resulting from the diagonalization of the above matrix can be written as
\begin{equation}
m^2_{\widetilde{\varphi}_{1,2}}=\frac{1}{4} (\mathcal{A} _1\mp\mathcal{F}_1 \mathcal{B}_1),
\end{equation}
with
\begin{equation}
\mathcal{A}_1= v^2(\lambda _3 +\lambda _4 +\lambda _8 )+(\lambda _9 +\lambda _{17}) v_2^2+(\lambda _{10} +\lambda _{18})v_6^2+(\lambda _{11}+\lambda _{19}) v_8^2-2( \lambda _{24} v_2 v_8+ v_6 \mu_s -\mu^2_3- \mu^2_4 ) ,
\end{equation}
\begin{align}
\mathcal{B}_1=&\bigg\{v^4 (\lambda_3 + \lambda_4 + \lambda_8)^2 + 2 v^2 v_8^2 (\lambda_3 + \lambda_4 + \lambda_8) (\lambda_{11} + \lambda_{19})\nonumber\\
&+ 2 v^2 v_2 (\lambda_3 + \lambda_4 + \lambda_8) \left[v_2 (\lambda_9 + \lambda_{17}) - 2 v_8 \lambda_{24}\right]
+ v^2 v_8^2 (\lambda_{11} + \lambda_{19})^2 + 2 v^2 v_8 \lambda_{11} \left[v_2 (\lambda_9 + \lambda_{17}) - 2 v_8 \lambda_{24}\right]\nonumber
\\
&+ 4 v^2 v_2^2 \lambda_8 \lambda_9 + 4 v^2 v_6^2 \lambda_8 \lambda_{10} + 4 v^2 v_8^2 \lambda_8 \lambda_{11}
+ 2 v^2 v_6^2 \lambda_9 \lambda_{18} + 2 v^2 v_8^2 \lambda_9 \lambda_{19} 
+ 2 v^2 v_6^2 \lambda_{10} \lambda_{17}+ 2 v^2 v_6^2 \lambda_{11} \lambda_{18}\nonumber\\
& + 2 v^2 v_8^2 \lambda_{11} \lambda_{17} + v_2^4 \lambda_9 \lambda_{17}
+ v_6^4 \lambda_{10} \lambda_{18} + v_8^4 \lambda_{11} \lambda_{19}
+ 4 v_2^2 v_6^2 \lambda_{10} \lambda_{17} + 2 v_6^2 v_8^2 \lambda_{10} \lambda_{19}
+ 2 v_2^2 v_8^2 \lambda_{11} \lambda_{17} \nonumber\\
&+ 2 v_6^2 v_8^2 \lambda_{11} \lambda_{18}
- 4 v_2^3 v_8 \lambda_9 \lambda_{24} - 4 v_2 v_6^2 v_8 \lambda_{10} \lambda_{24} - 4 v_2 v_8^3 \lambda_{11} \lambda_{24}
- 2 v^2 v_6^2 \lambda_{25}^2 - 4 v_2^2 v_6 \lambda_9 \mu_s\nonumber
\\
&- 4 v_6^3 \lambda_{10} \mu_s - 4 v_6 v_8^2 \lambda_{11} \mu_s + 2 v^2 \lambda_3 (v^2 \lambda_8 + v_2^2 \lambda_{17}
+ v_6^2 \lambda_{18} + v_8^2 \lambda_{19} - 2 v_2 v_8 \lambda_{24} - 2 v_6 \mu_s)\nonumber
\\
&+ 2 v^2 \lambda_4 (v^2 \lambda_8 + v_2^2 \lambda_{17} + v_6^2 \lambda_{18}
+ v_8^2 \lambda_{19} - 2 v_2 v_8 \lambda_{24} - 2 v_6 \mu_s) + 4 v^2 v_6 \lambda_{25} \mu_t - 2 v^2 \mu_t^2\bigg\}^{1/2},
\end{align}
and
\begin{equation}
\mathcal{F}_1=\text{sign}\left[v^2(\lambda_3+\lambda_4-\lambda_8)+v_8^2(\lambda_{11}-\lambda_{19})+v_2\left(\lambda_9 v_2-\lambda_{17}v_2+2\lambda_{24}v_8\right)+v_6\left(\lambda_{10}v_6-\lambda_{18}v_6+2\mu_s\right)+2(\mu_3^2-\mu_4^2)\right].
\end{equation}

After SSB, the scalar $G_1=a_1$ remains massless and can be identified with a Goldstone boson eaten by a linear combination of the neutral gauge fields.  The remaining states in the basis  $(a_2,a_6,a_8)$ have a squared mass matrix of the form 
\begin{equation}
M_2^2=\frac{1}{2}\left(
\begin{array}{ccc}
 \frac{v_6 \left(v_8 \mu_u-9 \lambda_{23} v_2^2 \right)}{ v_2} &  3 \lambda _{23} v_2^2+v_8 \mu _u  & - v_6 \mu _u \\
  3 \lambda _{23} v_2^2+v_8 \mu _u & \frac{v_2 v_8 \mu _u-\lambda _{23} v_2^3}{ v_6} & - v_2 \mu _u \\
 - v_6 \mu _u & - v_2 \mu _u & \frac{v_2 v_6 \mu _u}{ v_8} \\
\end{array}
\right).
\end{equation}
This matrix has rank 2, and therefore one of the physical states obtained from its diagonalization is the second Goldstone boson needed to give masses to the two neutral gauge bosons of the model, identified as
\begin{equation}
G_2=\frac{v_2a_2+3v_6 a_6+4v_8 a_8}{\sqrt{v_2^2+9v_6^2+16v_8^2}}.
\end{equation} 
From the non-zero eigenvalues, we can conclude that the physical CP-odd scalar spectrum  is completed by two real scalars $A_1$ and $A_2$ with squared masses
\begin{equation}
\begin{split}
m^2_{A_{1,2}} =& \frac{1}{4 v_2 v_6 v_8}\bigg\{ \lambda _{23} v_2^2 \left(v_2^2+9 v_6^2\right) v_8-\left[\left(v_6^2+v_8^2\right) v_2^2+v_6^2 v_8^2\right] \mu _u \\
&\mp \sqrt{4 \lambda _{23} v_6^2 v_8 \left(v_2^2+9 v_6^2+16 v_8^2\right) v_2^4 \mu _u+\left\lbrace \lambda _{23} v_2^2 \left(v_2^2+9 v_6^2\right) v_8-\left[\left(v_6^2+v_8^2\right) v_2^2+v_6^2 v_8^2\right] \mu _u \right\rbrace ^2}\bigg\}.
\end{split}
\end{equation}

\subsection{CP-even sector}
Regarding the neutral  CP-even sector, the dark scalars  in the basis  $(s_1^{\prime},s_2^{\prime})$  mix through the  squared mass matrix
\begin{equation}
M_3^2=\frac{1}{2}\left(
\begin{array}{cc}
  \left(\lambda_3+\lambda _4\right) v^2+\lambda _{9}v_{2}^2+\lambda _{10} v_{6}^2+\lambda _{11} v_{8}^2+2 \mu^2_3 &  v \left(\lambda _{25} v_6+\mu_t\right) \\
 v \left(\lambda _{25} v_6+\mu_t\right) &  \lambda _8 v^2+\lambda _{17} v_2^2+\lambda_{18} v_6^2+\lambda _{19} v_8^2+2 \lambda _{24} v_2 v_8+2 v_6 \mu_s+2\mu^2_4 \\
\end{array}
\right)\, ,
\label{eq:sp-matrix}
\end{equation}\\
which is diagonalized by
\begin{equation}
  \left(\begin{array}{c}
    \varphi_1\\
    \varphi_2
  \end{array}
  \right)
  =U^s
  \left(\begin{array}{c}
    s_1'\\
    s_2'
  \end{array}\right)
  =\left(\begin{array}{cc}
    \cos\theta_s& \sin\theta_s\\
    -\sin\theta_s& \cos\theta_s
  \end{array}\right)
  \left(\begin{array}{c}
    s_1'\\
    s_2'
  \end{array}\right)\, ,
\label{eq:even_orthogonal}
\end{equation}
where the mixing angle is determined by
\begin{equation}
\tan2\theta_s=\frac{2v(\mu_t+\lambda_{25}v_6)}{v^2(\lambda_3+\lambda_4-\lambda_8)+v_8^2(\lambda_{11}-\lambda_{19})+v_2(\lambda_9v_2-\lambda_{17}v_2-2\lambda_{24}v_8)+v_6(\lambda_{10}v_6-\lambda_{18}v_6-2\mu_s)+2(\mu_3^2-\mu_4^2)}\, .
\label{eq:t2beta}
\end{equation}
The mass eigenvalues of the real physical scalars $\varphi_1$ and  $\varphi_2$ that diagonalize the above matrix can be written as
\begin{equation}
m^2_{{\varphi}_{1,2}}=\frac{1}{4} (\mathcal{A} _2\mp \mathcal{F}_2\mathcal{B}_2),
\end{equation}
with 
\begin{equation}
\mathcal{A}_2= v^2(\lambda _3 +\lambda _4 +\lambda _8 )+(\lambda _9 +\lambda _{17}) v_2^2+(\lambda _{10} +\lambda _{18})v_6^2+(\lambda _{11}+\lambda _{19}) v_8^2+2( \lambda _{24} v_2 v_8+ v_6 \mu_s +\mu^2_3+ \mu^2_4 )\, ,
\end{equation}
\begin{align}
\mathcal{B}_2=&\bigg\{v^4 (\lambda_3 + \lambda_4 + \lambda_8)^2 + 2 v^2 v_8^2 (\lambda_3 + \lambda_4 + \lambda_8) (\lambda_{11} + \lambda_{19})\nonumber\\
&+ 2 v^2 v_2 (\lambda_3 + \lambda_4 + \lambda_8) \left[v_2 (\lambda_9 + \lambda_{17}) +2 v_8 \lambda_{24}\right]
+ v^2 v_8^2 (\lambda_{11} + \lambda_{19})^2 + 2 v^2 v_8 \lambda_{11} \left[v_2 (\lambda_9 + \lambda_{17}) + 2 v_8 \lambda_{24}\right]\nonumber
\\
&+ 4 v^2 v_2^2 \lambda_8 \lambda_9 + 4 v^2 v_6^2 \lambda_8 \lambda_{10} + 4 v^2 v_8^2 \lambda_8 \lambda_{11}
+ 2 v^2 v_6^2 \lambda_9 \lambda_{18} + 2 v^2 v_8^2 \lambda_9 \lambda_{19} 
+ 2 v^2 v_6^2 \lambda_{10} \lambda_{17}+ 2 v^2 v_6^2 \lambda_{11} \lambda_{18}\nonumber\\
& + 2 v^2 v_8^2 \lambda_{11} \lambda_{17} + v_2^4 \lambda_9 \lambda_{17}
+ v_6^4 \lambda_{10} \lambda_{18} + v_8^4 \lambda_{11} \lambda_{19}
+ 4 v_2^2 v_6^2 \lambda_{10} \lambda_{17} + 2 v_6^2 v_8^2 \lambda_{10} \lambda_{19}
+ 2 v_2^2 v_8^2 \lambda_{11} \lambda_{17} \nonumber\\
&+ 2 v_6^2 v_8^2 \lambda_{11} \lambda_{18}
+ 4 v_2^3 v_8 \lambda_9 \lambda_{24} + 4 v_2 v_6^2 v_8 \lambda_{10} \lambda_{24} + 4 v_2 v_8^3 \lambda_{11} \lambda_{24}
-2 v^2 v_6^2 \lambda_{25}^2 +4 v_2^2 v_6 \lambda_9 \mu_s\nonumber
\\
&+4 v_6^3 \lambda_{10} \mu_s +4 v_6 v_8^2 \lambda_{11} \mu_s + 2 v^2 \lambda_3 (v^2 \lambda_8 + v_2^2 \lambda_{17}
+ v_6^2 \lambda_{18} + v_8^2 \lambda_{19} + 2 v_2 v_8 \lambda_{24} + 2 v_6 \mu_s)\nonumber
\\
&+ 2 v^2 \lambda_4 (v^2 \lambda_8 + v_2^2 \lambda_{17} + v_6^2 \lambda_{18}
+ v_8^2 \lambda_{19} + 2 v_2 v_8 \lambda_{24} + 2 v_6 \mu_s) - 4 v^2 v_6 \lambda_{25} \mu_t - 2 v^2 \mu_t^2\bigg\}^{1/2},
\end{align}
and
\begin{equation}
\mathcal{F}_2=\text{sign}\left[v^2(\lambda_3+\lambda_4-\lambda_8)+v_8^2(\lambda_{11}-\lambda_{19})+v_2(\lambda_9v_2-\lambda_{17}v_2-2\lambda_{24}v_8)+v_6(\lambda_{10}v_6-\lambda_{18}v_6-2\mu_s)+2(\mu_3^2-\mu_4^2)\right].
\end{equation}

The remaining scalars in the basis $(s_1,s_2,s_6,s_8)$ have a squared mass matrix given by
 \begin{equation}
 M_4^2=
 \begin{pmatrix}
 2 v^2 \lambda_1 & v v_2 \lambda_5 & v v_6 \lambda_6 & v v_8 \lambda_7 \\
 v v_2 \lambda_5 & 2 v_2^2 \lambda_{13} + \frac{3}{2} v_2 v_6 \lambda_{23} + \frac{v_6 v_8 \mu_u}{2 v_2} & v_2 v_6 \lambda_{20} + \frac{3 v_2^2 \lambda_{23}}{2} - \frac{v_8 \mu_u}{2} & v_2 v_8 \lambda_{21} - \frac{v_6 \mu_u}{2} \\
 v v_6 \lambda_6 & v_2 v_6 \lambda_{20} + \frac{3 v_2^2 \lambda_{23}}{2} - \frac{v_8 \mu_u}{2} & 2 v_6^2 \lambda_{14} - \frac{1}{2} \frac{v_2^3}{v_6} \lambda_{23} + \frac{1}{2} v_2 v_8 \mu_u & v_6 v_8 \lambda_{22} - \frac{v_2 \mu_u}{2} \\
 v v_8 \lambda_7 & v_2 v_8 \lambda_{21} - \frac{v_6 \mu_u}{2} & v_6 v_8 \lambda_{22} - \frac{v_2 \mu_u}{2} & 2 v_8^2 \lambda_{15} + \frac{v_2 v_6 \mu_u}{2 v_8}
\end{pmatrix}\, .
 \end{equation}
Assuming that  the  hierarchy $v_8, v_6, v_2, \mu_u\gg v$ holds, it is convenient to write $M_4^2$ as 
\begin{equation}
M_4^2=
\begin{pmatrix}
A&B\\
B^T&D
\end{pmatrix}\,,
\end{equation}
where the blocks have the explicit form
\begin{align}
A&=2v^2\lambda_1\, ,\\
B&=
\begin{pmatrix}
\lambda_5 v v_2 &
\lambda_6 v v_{6}&
\lambda_7 v v_8
\end{pmatrix}\,,\\
D&=
\begin{pmatrix}
    2 v_2^2 \lambda_{13} + \frac{3}{2} v_2 v_6 \lambda_{23} + \frac{v_6 v_8 \mu_u}{2 v_2} & v_2 v_6 \lambda_{20} + \frac{3 v_2^2 \lambda_{23}}{2} - \frac{v_8 \mu_u}{2} & v_2 v_8 \lambda_{21} - \frac{v_6 \mu_u}{2} \\
    v_2 v_6 \lambda_{20} + \frac{3 v_2^2 \lambda_{23}}{2} - \frac{v_8 \mu_u}{2} & \frac{4 v_6^3 \lambda_{14} - v_2^3 \lambda_{23} + v_2 v_8 \mu_u}{2 v_6} & v_6 v_8 \lambda_{22} - \frac{v_2 \mu_u}{2} \\
    v_2 v_8 \lambda_{21} - \frac{v_6 \mu_u}{2} & v_6 v_8 \lambda_{22} - \frac{v_2 \mu_u}{2} & 2 v_8^2 \lambda_{15} + \frac{v_2 v_6 \mu_u}{2 v_8}
\end{pmatrix}\,.
\end{align}
These blocks have characteristic scales that inherit the hierarchy $|A|<|B| <|D | $. It is possible to perform a block diagonalization, through a modal block matrix $U_4$ yielding
\begin{equation}
U_4^{-1}M_{4}^2U_4\approx
\begin{pmatrix}
m_h^2& 0\\
0 & M^{'2}_{4}
\end{pmatrix}\,,
\end{equation}
with
\begin{align}
m^2_h\approx A=2v^2\lambda_1,	& 	\qquad	&M^{'2}_{4}\approx D-B\cdot A^{-1}\cdot B^T.
\end{align}
The lightest eigenvalue  corresponds to the Standard Model Higgs field, mostly composed by the $s_1$ scalar. Defining the auxiliary  couplings  $\lambda'_{13}=2\lambda_{13}-\frac{\lambda_5^2}{2\lambda_1}, \lambda'_{20}=\lambda_{20}-\frac{\lambda_5 \lambda_6}{2\lambda_1}, \lambda'_{21}=\lambda_{21}-\frac{\lambda_5 \lambda_7}{2\lambda_1}, \lambda'_{14}=2\lambda_{14}-\frac{\lambda_6^2}{2\lambda_1},\lambda'_{22}=\lambda_{22}-\frac{\lambda_6\lambda_7}{2\lambda_1}, \lambda'_{15}=2\lambda_{15}-\frac{\lambda_7^2}{2\lambda_1}$, the  $M^{'2}_{4}$ block can be written as
\begin{equation}
M^{'2}_{4}\approx
\begin{pmatrix}
v_2^2 \lambda'_{13} + \frac{3}{2} v_2 v_6 \lambda_{23} + \frac{v_6 v_8 \mu_u}{2 v_2} & v_2 v_6 \lambda'_{20} + \frac{3 v_2^2 \lambda_{23}}{2} - \frac{v_8 \mu_u}{2} & v_2 v_8 \lambda'_{21} - \frac{v_6 \mu_u}{2} \\
v_2 v_6 \lambda'_{20} + \frac{3 v_2^2 \lambda_{23}}{2} - \frac{v_8 \mu_u}{2} & \lambda'_{14} v_6^2 - \frac{1}{2} \frac{v_2^3}{v_6} \lambda_{23} + \frac{1}{2} v_2 v_8 \mu_u & v_6 v_8 \lambda'_{22} - \frac{v_2 \mu_u}{2} \\
v_2 v_8 \lambda'_{21} - \frac{v_6 \mu_u}{2} & v_6 v_8 \lambda'_{22} - \frac{v_2 \mu_u}{2} & v_8^2 \lambda'_{15} + \frac{v_2 v_6 \mu_u}{2 v_8}
\end{pmatrix}\, .
\end{equation}
This matrix has rank 3 for arbitrary values of its parameters, and therefore, there are three physical CP-even neutral scalars $H_a$, $a=1,2,3$, with masses of the same order as the $B-L$ breaking scale, composed mostly by the $s_2$,$s_6$ and $s_8$ scalars.

\subsection{Gauge Bosons}
We define the $U(1)_Y$ and $U(1)_{B-L}$ gauge fields as $B_\mu$ and $B_\mu^{'}$, respectively. Thus, the covariant derivative has the form
\begin{equation}
\mathcal{D}_{\mu}\equiv I\cdot\partial_\mu+igT^a W_\mu^a+ig_1 Y B_\mu+i(\tilde{g}Y+g'_1 Y_{B-L})B'_\mu\, ,
\end{equation}
where $g$, $g_1$ and $g'_1$ are the $\mathrm{SU(2)}_W$, $\mathrm{U(1)}_Y$ and $\mathrm{U(1)}_{B-L}$ coupling constants, respectively. Here $\tilde{g}$ parameterizes the kinetic mixing between the hypercharge gauge boson $B_\mu$ and the new one $B'_\mu$ associated to the $U(1)_{B-L}$ symmetry \footnote{For a suitable choice of $g'_1$ and $\tilde{g}$, it is possible to obtain a variety of  different $U(1)_X$ extensions to the SM. These parameters control the coupling structure and the mixing among the neutral gauge eigenstates $(B_\mu, W^3_\mu, B'_\mu)$ prior to mass diagonalization. 
For instance, $g'_1=g_1$ and $\tilde{g}=0$ at EW scale corresponds to the sequential SM. Similarly, taking $\tilde{g}=-2g'_1$ at EW scale produces the $U(1)_R$ model in \cite{delAguila:1988jz,delAguila:1995rb}. In contrast, the pure $B-L$ model is defined by $U(1)_X = U(1)_{B-L}$ and $\tilde{g} = 0$, which implies that the field $B'_\mu$ does not mix at tree level with the SM neutral gauge bosons and becomes directly the mass eigenstate $Z'_\mu$.}
The squared mass matrix corresponding to the neutral gauge sector in the basis $(B_\mu, W_{\mu}^3, B'_\mu)$ is given by
\begin{equation}
M_{G}^2=
\begin{pmatrix}
\frac{1}{4} g_{1}^2 v^2 & -\frac{1}{4} g_1 g v^2 & \frac{1}{4} g_1 \tilde{g} v^2 \\
-\frac{1}{4} g_1 g v^2 & \frac{1}{4} g^2 v^2 & -\frac{1}{4} g \tilde{g} v^2 \\
\frac{1}{4} g_1 \tilde{g} v^2 & -\frac{1}{4} g \tilde{g} v^2 & \frac{\tilde{g}^2}{4} v^2 + g_1^{'2} (4 v_2^2 + 36 v_6^2 + 64 v_8^2)
\end{pmatrix}\, .
\end{equation}
We are interested in
the `pure' $B-L$ model, which is obtained by setting the condition $\tilde{g}(Q_{EW})=0$\footnote{In models with non-zero kinetic mixing parameter $\tilde{g}$, the physical states $Z$ and $Z'$ can contain components of all three gauge eigenstates, and may thus lead to observable interference effects. However, in the pure $B-L$ model defined by $\tilde{g} = 0$, no such mixing occurs: $B'_\mu$ decouples at tree level from the Standard Model neutral gauge bosons, and one can identify $B'_\mu = Z'_\mu$ directly.},  implying that the $B'_\mu$ field does not mix with the SM gauge bosons at tree level, at electroweak scale $Q_{EW}$.
This matrix can be diagonalized through an orthogonal transformation as
\begin{equation}\label{mixing}
\begin{pmatrix}
A_\mu\\
Z_\mu\\
Z'_\mu
\end{pmatrix}
=
\begin{pmatrix}
c_W & s_W & 0\\
-s_W c_\alpha & c_W c_\alpha & -c_W s_\alpha\\
s_W s_\alpha & -c_W s_\alpha & c_\alpha
\end{pmatrix}
\begin{pmatrix}
B_\mu\\
W_{\mu}^3\\
B'_\mu
\end{pmatrix}\, ,
\end{equation}
with $\theta_W$ as the Weinberg angle defined by $\tan\theta_W\equiv t_W=s_W/c_W=g_1/g$. The gauge boson masses are
\begin{align}
  M_A&=0\, ,\\
  M_Z^2&=\frac{g_1 v^2(g_1-\tilde{g}s_W t_\alpha)}{4s_W^2}\, , & M_{Z'}^2&=\frac{g_1 v^2(g_1+\tilde{g}s_W/t_\alpha)}{4s_W^2}\, , 
\end{align}
where the mixing angle $\alpha$ satisfies
\begin{align}
    \sin 2\alpha &=\frac{2 \tilde{g} v^2 \sqrt{g^2+g_1^2}}{\sqrt{\left(v^2 \left(g^2+g_1^2-\tilde{g}^2\right)-16 g_1^{'2} \left(v_2^2+9 v_6^2+16 v_8^2\right)\right)^2+4 \tilde{g}^2 v^4 \left(g^2+g_1^2\right)}}\, ,
\end{align}
and is constrained mainly by the LEP, CMS and ATLAS collaborations \cite{delAguila:2010mx}, \cite{Osland:2022ryb}, \cite{Osland:2020onj} as $\left|\alpha\right|\leq 10^{-3}\,$.
In the pure $B-L$ scenario, setting $\tilde{g}= 0$ at EW scale leads to a vanishing $\alpha$, and the neutral gauge boson  masses reduce to
\begin{align}
  M_Z&\approx \frac{g_1 v}{2s_W}, & M_{Z'}&=2g_1' x\, ,
\end{align}
with $x=\sqrt{v_2^2+9v_6^2+16v_8^2}$.
%%%%%%%%%%%%%%%%%%%%%%%%%%%%%%%%%%%%%%
\section{Neutrino masses}
\label{sec:Nu mass}
\begin{figure}[htbp]
\begin{center}
\includegraphics[scale=1.5]{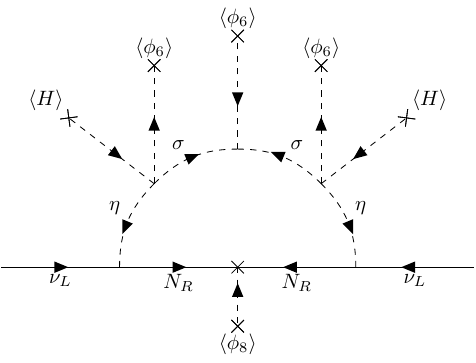}
\caption{One-loop scotogenic neutrino mass generation mechanism.}
\label{Neutrino_diagram}\label{scot}
\end{center}
\end{figure}

The Yukawa Lagrangian of the model is 
\begin{equation}\label{eqn:Yuk}
\begin{split}
-L_Y=&y^e_{ab} \overline{L}_{aL} H e_{bR}+y^u_{ab} \overline{Q}_{aL} \widetilde{H} u_{bR}+y^d_{ab} \overline{Q}_{aL}H d_{bR}+y^{M}_{ij} \overline{N^c}_{i R}\phi_8 N_{j R} +y^{N}_{ai}\overline{L}_{aL} \widetilde{\eta} N_{i R} \\
&+  y^{X}_1\overline{F}_L \phi^{\star}_2 X_{R}+ y^{X}_2\overline{F^c}_R \phi^{\star}_8 X_{R}+ y^{F}_1\overline{F}_L \phi_6 F^c_{L}+ y^{F}_2\overline{F^c}_R \phi^{\star}_6 F_{R}+m^{D}_F \overline{F}_L F_R+\mathrm{h.c.}\, ,
\end{split}
\end{equation}
with $\widetilde{H}\equiv i\tau_2H^{\star}$, $a,b=1,2,3$, and $i,j=1,2$. The relevant interactions for the active neutrinos mass generation mechanism are
\begin{equation}
-L_Y \supset y^M_{ij} \overline{N^c}_{iR}\phi_8 N_{jR}+y^{N}_{ai}\overline{L}_{aL} \widetilde{\eta} N_{i R}.
\label{eqn:Y_mass}
\end{equation} 
We can write the above terms 
in the physical basis for the mediators, through the following transformation 
\begin{equation} 
N_R=US_R,
 \end{equation}
where $U$ is the unitary matrix that diagonalizes the mass matrix  $M_M=y^Mv_8$,
according to $M^\prime =U^T M_M U= \mathrm{diag} (M_{1},M_{2})$.
Then, from Eq.(\ref{eqn:Y_mass}) we have
\begin{equation}
-L_Y \supset \frac{(y^N U)_{ai}U_{j1}^s}{\sqrt{2}} \overline{\nu}_{aL}\varphi_j S_{iR}+i \frac{(y^N U)_{ai}U_{j1}^a}{\sqrt{2}} \overline{\nu}_{aL} \widetilde{\varphi}_j S_{i R}+\sum_{k =1}^2 \frac{M_{k}}{2}\overline{S^c}_{k R} \phi_8 S_{k R}.
\end{equation}
The light active neutrino masses are generated at one-loop level by the diagram shown in Fig.(\ref{scot}).
The resulting light neutrino mass matrix can be written as
\begin{equation}
(M_{\nu})_{ab}=\sum_{k=1}^2 \sum_{j=1}^2 \frac{(y^N U)_{ak} (y^N U)_{b k}M_{k}}{32 \pi ^2}\left[(U_{j1}^s)^2 \frac{m_{\varphi_j}^2}{m_{\varphi_j}^2 - M_{k}^2} \log\left(\frac{m_{\varphi_j}^2}{M_{k}^2}\right)-(U_{j1}^a)^2 \frac{m_{\widetilde{\varphi}_j}^2}{m_{\widetilde{\varphi}_j}^2 - M_{k}^2} \log\left(\frac{m_{\widetilde{\varphi}_j}^2}{M_{k}^2}\right) \right] ,\label{eqn:nu_mass}
\end{equation}
where $U_{11}^{a,s}$, $U_{21}^{s,a}$ correspond to the $(1,1)$ and $(2,1)$ components of the orthogonal matrices $U^a$ and $U^s$ according to Eqs.(\ref{eq:odd_orthogonal},~\ref{eq:even_orthogonal}) for the mixing angles $\theta_a, \theta_s$ in Eqs.(\ref{eq:t2delta},~\ref{eq:t2beta}). Breaking down the individual $j=1,2$, we get
\begin{align}
  (M_{\nu})_{ab}=\sum_{k=1}^2\frac{(y^N U)_{ak}(y^N U)_{bk}M_k}{32\pi^2}&\left[\cos^2\theta_s\frac{m_{\varphi_1}^2}{m_{\varphi_1}^2-M_k^2}\log\left(\frac{m_{\varphi_1}^2}{M_k^2}\right)+\sin^2\theta_a\frac{m_{\varphi_2}^2}{m_{\varphi_2}^2-M_k^2}\log\left(\frac{m_{\varphi_2}^2}{M_k^2}\right)\right.\nonumber\\
  &\left.-\cos^2\theta_a\frac{m_{\widetilde{\varphi}_1}}{m_{\widetilde{\varphi}_1^2}-M_k^2}\log\left(\frac{m_{\widetilde{\varphi}_1}^2}{M_k^2}\right)-\sin^2\theta_s\frac{m_{\widetilde{\varphi}_2}^2}{m_{\widetilde{\varphi}_2^2}-M_k^2}\log\left(\frac{m_{\widetilde{\varphi}_2^2}}{M_k^2}\right)\right]\, .
\end{align}
 Notice that the crucial parameters in the scalar potential responsible for the emergence of the active neutrino mass matrix are the quartic couplings  $\lambda_{24}, \lambda_{25}$ and the cubic coupling $\mu_s$, contained in the mixing angles above. In fact the matrices $M_1^2$ and $M_3^2$ in Eqs.(\ref{eq:ap-matrix},\ref{eq:sp-matrix}) become degenerate in the limit $\lambda_{24}, \lambda_{25}, \mu_s \to 0$, and so do their respective mixing angles $\theta_a, \theta_s$, and their squared mass eigenvalues $m^2_{\widetilde{\varphi}_{1,2}}$, $m^2_{\varphi_{1,2}}$, leading to a vanishing active neutrino mass matrix and consequently, to lepton number restoration in the active neutrino sector.

After SSB, the remaining electrically neutral fermions in Eq.(\ref{eqn:Yuk}) mix in the basis $(X,F^c_L,F_R)$ through the mass matrix
\begin{equation}
M_{XF}=\frac{1}{\sqrt{2}}\begin{pmatrix}
0 & y_1^Xv_2&y_2^Xv_2 \\
y_1^Xv_2 & y_1^Fv_6 & \sqrt{2} m_F^D \\
y_2^Xv_2 &  \sqrt{2} m_F^D & y_2^Fv_6 
\end{pmatrix}.
\end{equation}
This matrix has rank 3, and after diagonalization renders three physical majorana fermions with masses at the scale of $B-L$ symmetry breaking.

%%%%%%%%%%%%%%%%%%%%%%%%%%%%%%%%%%%%
\section{Neutrinoless double beta decay}
\label{sec:Neutrinoless}
%%%%%%%%%%%%%%%%%%%%%%%%%%%%%%%%%%%%

The observation of neutrinoless double beta ($0\nu\beta\beta$) decay can undoubtedly establish the majorana nature of neutrinos \cite{Schechter:1981bd}, as predicted in our model.  The effective Majorana mass involved in the standard light neutrino-mediated amplitude of the process is given by
\begin{equation}
\left\langle m_{\beta \beta}\right\rangle = |\cos^2{\theta_{12}}\cos^2{\theta_{13}}m_1+\sin^2{\theta_{12}}\cos^2{\theta_{13}}m_2 e^{2i\phi_{12}}+\sin^2{\theta_{13}}m_3 e^{2i\phi_{13}}| \, ,
\label{eq:nuless_efectivemass}
\end{equation}
where we have adopted the symmetric parameterization of the lepton mixing matrix, introduced in \cite{Schechter:1980gr} and revisited in \cite{Rodejohann:2011vc}. This parameterization is particularly useful in the analysis of $0\nu\beta\beta$ decay since the three independent phases physical phases  $\phi_{12}$, $\phi_{13}$ and   $\phi_{23}$ are related to the Dirac CP violating phase according to
\begin{equation}
\delta=\phi_{13}-\phi_{12}-\phi_{23},
\end{equation}
and only two of them appear explicitly in the  $0\nu\beta\beta$ decay amplitude in Eq.(\ref{eq:nuless_efectivemass}).

\begin{figure}[htbp]
\begin{center}
\begin{subfigure}[b]{0.45\linewidth} 
\includegraphics[width=\linewidth]{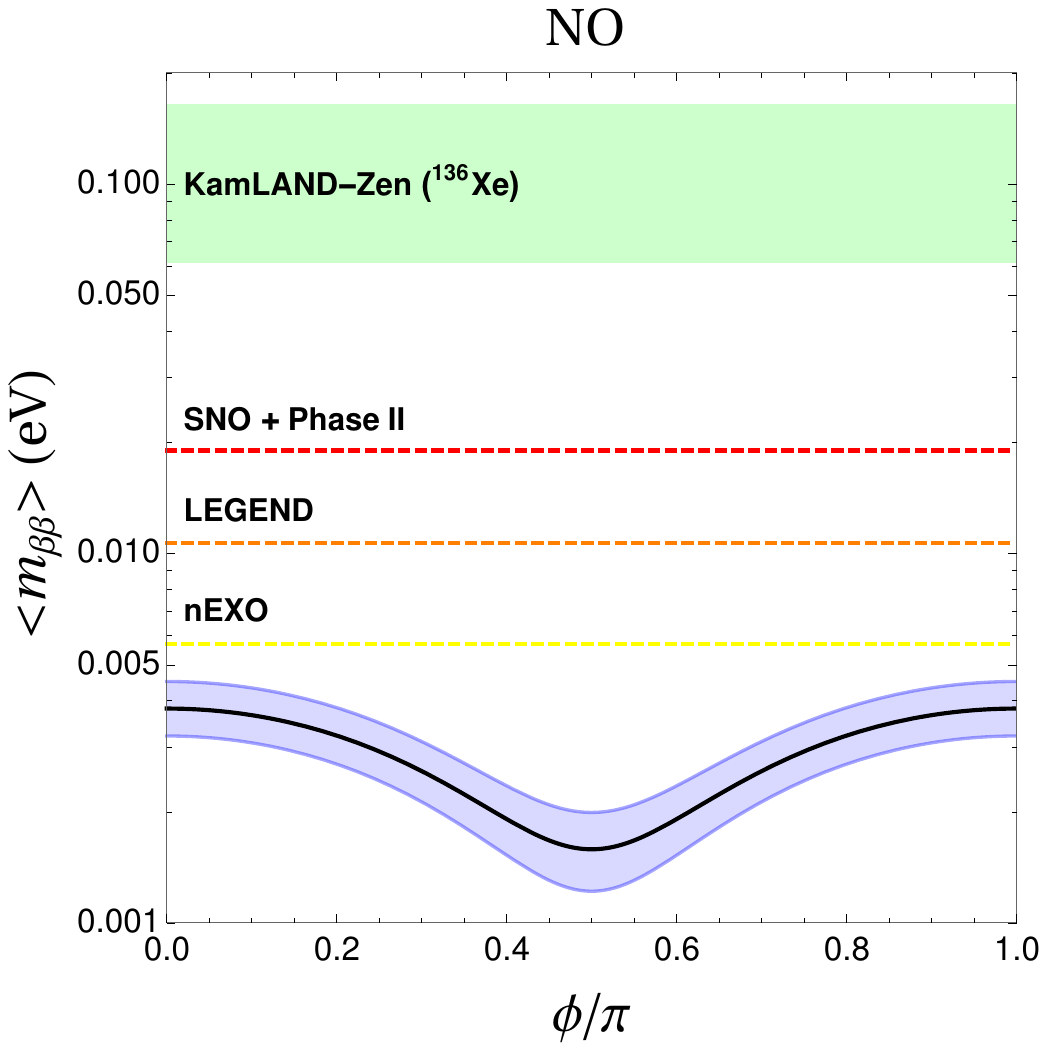}
\caption{Normal Ordering (NO)}
\label{fig:2a}
\end{subfigure}
\begin{subfigure}[b]{0.45\linewidth} 
\includegraphics[width=\linewidth]{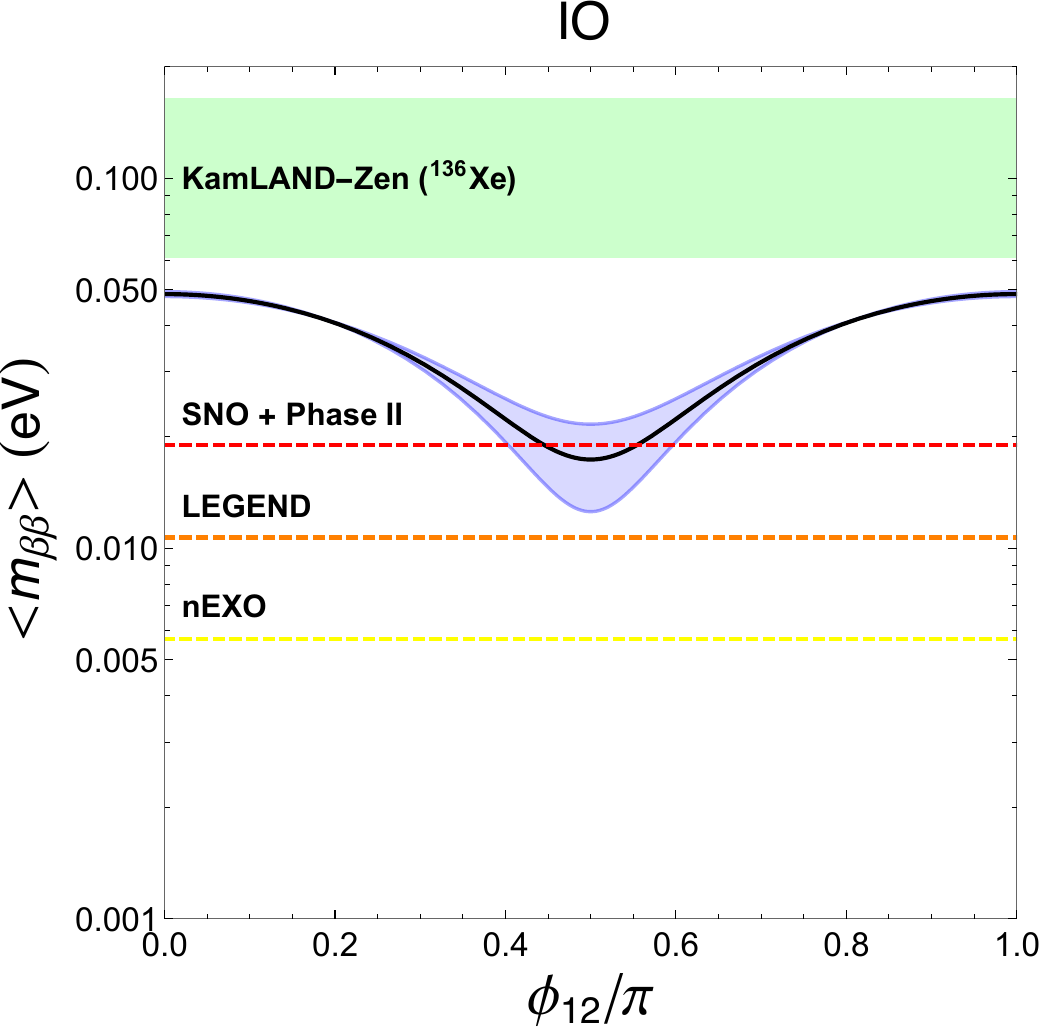}
\caption{Inverted Ordering (IO)}
\label{fig:2b}
\end{subfigure}
\caption{Effective Majorana mass  for the case of normal (a) and inverted (b) mass ordering
 vs relative Majorana phase $\phi$ and $\phi_{12}$, respectively. The blue contours represent the allowed $3\sigma$ values consistent with the global fit of oscillation parameters in \cite{deSalas:2020pgw}. Horizontal bands in both figures represent current experimental limits
 and future sensitivities.}\label{fig:2}
\end{center}
\end{figure}

Since the light active neutrino mass matrix in Eq.(\ref{eqn:nu_mass}) has rank two, the lightest neutrino eigenstate is massless. 
For normal ordering (NO), we have $m_1=0$ and therefore \eqref{eq:nuless_efectivemass} reduces to 
\begin{equation}
\left\langle m_{\beta \beta}\right\rangle_{NO}=  |\sin^2{\theta_{12}}\cos^2{\theta_{13}}m_2 e^{2i\phi_{12}}+\sin^2{\theta_{13}}m_3 e^{2i\phi_{13}}| \, ,
\label{eq:nuless_efectivemassNO_0}
\end{equation}
or, equivalently
\begin{equation}
\left\langle m_{\beta \beta}\right\rangle_{NO}=  |\sin^2{\theta_{12}}\cos^2{\theta_{13}}m_2 e^{2i\phi}+\sin^2{\theta_{13}}m_3 | \, ,
\label{eq:nuless_efectivemassNO}
\end{equation}
where we have defined the relative Majorana phase
 the relative Majorana phase, defined by
\begin{equation}
\phi \equiv \phi_{12}-\phi_{13}.
\end{equation}
In the case of the inverted ordering (IO), the vanishing mass eigenvalue is $m_3$ and the corresponding effective mass becomes
\begin{equation}
\left\langle m_{\beta \beta}\right\rangle_{IO}=  |\cos^2{\theta_{12}}\cos^2{\theta_{13}}m_1+\sin^2{\theta_{12}}\cos^2{\theta_{13}}m_2 e^{2i\phi_{12}}| \, .
\label{eq:nuless_efectivemassIO}
\end{equation}
Notice that in the general expression for the amplitude in Eq.(\ref{eq:nuless_efectivemassNO_0}), destructive interference can lead to a vanishing effective mass in the NO scenario, currently preferred by oscillation data \cite{deSalas:2020pgw}. In the present model, the existence of a massless neutrino in the spectrum prevents the appearance of such destructive interference. This happens due to the fact that the amplitude depends essentially on one free parameter, the relative majorana phase $\phi$, as the remaining parameters are well constrained by oscillation experiments. In Fig.(\ref{fig:2}) we present the  effective majorana mass for the case of normal and inverted ordering, as functions of their relative Majorana phases, showing that the $0\nu\beta\beta$ amplitude never vanishes. Horizontal bands in both figures represent current experimental limits and future sensitivities. In particular, the green band represents the current experimental limits coming from: KamLAND - Zen ($^{136}\text{Xe}$) (61 - 165 meV)\cite{KamLAND-Zen:2016pfg}, the dashed  red, orange and yellow lines represent the lower experimental limits from  SNO + Phase II(19 meV)\cite{SNO:2015wyx} , LEGEND (10.7 meV) \cite{LEGEND:2017cdu} and nEXO (5.7  meV) \cite{nEXO:2017nam}, respectively. Remarkably, the resulting prediction for the effective mass in the case of IO lies within the expected sensitivity of the upcoming next generation experiments.

%%%%%%%%%%%%%%%%%%%%%%%%%%%%%%%%%%%%
\section{Dark Matter}
\label{sec:DM}
%%%%%%%%%%%%%%%%%%%%%%%%%%%%%%%%%%%%

The lightest electrically neutral  $M_P$-odd particle is automatically stable and becomes a natural WIMP DM candidate. Here, we study the scenario in which DM is identified with the lightest neutral scalar of the dark sector, assumed to be $\varphi_2$.  The phenomenology of a real scalar DM in our model was previously studied in a similar scotogenic scenario in \cite{CarcamoHernandez:2020ehn}. There, the viability of $\varphi_2$ as a WIMP DM was analyzed within a simplified scenario  where all the non-SM fields were assumed to be heavy and decouple, in the approximation $|\theta_s|\ll 1$ such that $\varphi_2$ is mostly composed by the real part of the electroweak singlet $\sigma$ and its coupling with the $Z$ boson is suppressed.
  In this approximation our DM candidate  annihilates mainly into a pair of Higgs fields, through the Higgs portal quartic scalar interaction described by the coupling $\lambda_8$ in the scalar potential in Eq.(\ref{Pot}), yielding viable relic densities, while evading direct detection bounds \cite{LZ:2022lsv}.
  This happens in two viable mass regions: near half of the Higgs mass, where resonant annihilation of dark matter into the Higgs boson takes place,
and also for masses $m_{\varphi_2}$ above $1\,\mathrm{TeV}$, where the direct detection constraints on the effective Higgs portal coupling are weak. 

\begin{figure}[htbp]
\begin{center}
\includegraphics[width=0.95\linewidth]{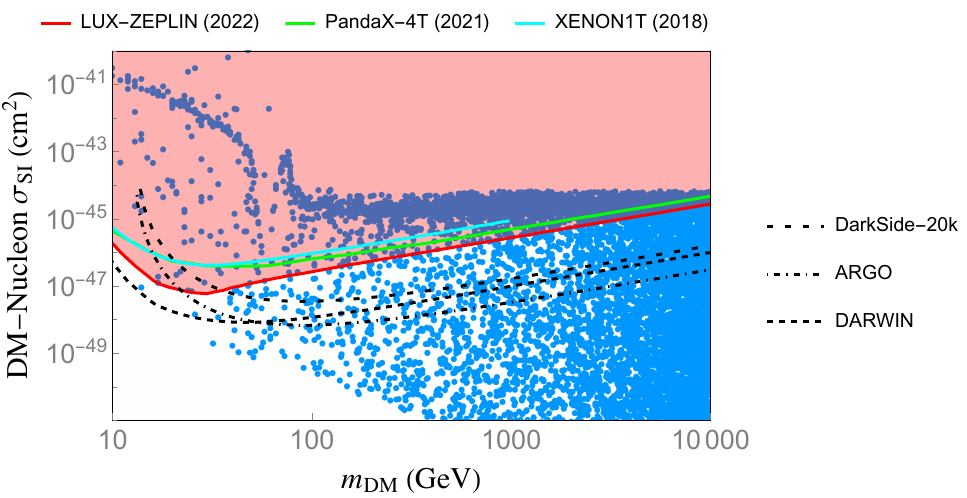}
\caption{Dark matter direct detection and relic abundance bounds. Each point represents a set of parameters that reproduces the correct relic abundance measured by PLANCK \cite{Planck:2018vyg}. Current limits from the LUX-ZEPLIN \cite{LZ:2022lsv}, PandaX-4T \cite{PandaX-4T:2021bab} and Xenon1T \cite{XENON:2018voc} experiments are shown, together with future experiment projected sensitivities \cite{Billard:2021uyg,DarkSide-20k:2017zyg,Schumann:2015cpa}.}
\label{fig:3}
\end{center}
\end{figure}

However, in this class of scotogenic models with physical scalar fields $\varphi_1$ and $\varphi_2$ composed by the real parts of the inert doublet $\eta$ and the scalar singlet $\sigma$, the parameter space for a viable WIMP DM is significatively widened by re-scattering effects if the masses of the physical scalars are almost degenerate $m_{\varphi_1}\approx m_{\varphi_2}$ \cite{Kakizaki:2016dza}.
The results of our analysis in this scenario are shown in Fig.(\ref{fig:3}), where we have varied randomly the relevant couplings in the range $0<|\lambda_3|,|\lambda_4|,|\lambda_8|<4\pi$ in order to ensure perturbativity, 
and we have scanned the mass parameters in $0<m_{\varphi_2}=m_{DM}<10^4\text{GeV}$, with $0<m_{\varphi_1}<1.2m_{\varphi_2}$ and the mixing angle within $0<|\theta_s|<0.01$. Each blue point corresponds to a set of parameters that reproduces the correct relic abundance $\Omega h=0.120\pm0.001$ \cite{Planck:2018vyg} through Higgs portal interactions.  As can be seen, there is a rather large parameter window below the current direct detection constraints, being those imposed by LUX-ZEPLIN the most restrictive ones \cite{LZ:2022lsv}.

%%%%%%%%%%%%%%%%%%%%%%%%%%%%%%%%%%%%
\section{Summary and conclusions}
\label{sec:Conclusions}
%%%%%%%%%%%%%%%%%%%%%%%%%%%%%%%%%%%%
In this work, we propose a new $\mathrm{SU}(3)\otimes \mathrm{SU}(2)_W\otimes\mathrm{U}(1)_Y\otimes \mathrm{U}(1)_{B-L}$ gauge model 
free of anomalies due to the introduction of three electrically neutral right-handed fermions with $B-L$ charge assignments $(-4,-4,5)$. The gauge symmetry is spontaneously broken by two units, 
resulting in a residual matter parity $M_P$. In this scenario two of the extra fermions become natural mediators 
for the radiative generation of light active neutrino masses through a 
scotogenic mechanism where the lightest $M_P$-odd neutral state can be identified as a WIMP DM candidate. We have studied the viability of the lightest scalar mediator in the scotogenic loop as the DM candidate, and we have found that there is a sizable region of parameter space where the model can reproduce the correct relic density, while evading the current direct detection constraints. We have also analyzed neutrinoless double beta decay, which has a lower bound in our model that comes from the fact that the light active neutrino mass matrix has rank two and therefore the lightest neutrino is predicted to be massless. During the revision of this work, the model in \cite{VanDong:2023thb} was brought to our attention. This model shares the same foundational concept and analogous outcomes with respect to ours, yet it features a different field content. This suggests the existence of an entire class of scotogenic models comprising two fermions with $B-L$ charge $-4$ functioning as intermediaries in the neutrino mass generation mechanism.
\black
\acknowledgements 
\noindent

Work supported by  CONAHCyT IxM project 749 and SNII 58928. The relic
abundance and direct detection constraints were calculated using the micrOMEGAS package \cite{Belanger:2018ccd} at GuaCAL (Guanajuato Computational Astroparticle Lab).

\bibliographystyle{utphys}
\bibliography{bibliography.bib}
\end{document}